\documentclass{aa} % 

\usepackage{enumerate}
\usepackage{graphicx}
\usepackage{color}  
\usepackage{txfonts}
\begin{document}

   \title{A statistical study of long--term evolution of coronal hole properties as observed by SDO}

   \author{S. G. Heinemann \inst{1}
          \and
          V. Jer\v{c}i\'{c} \inst{1,2}
          \and
          M. Temmer \inst{1}
          \and
          S. J. Hofmeister \inst{1}
          \and
          M. Dumbovi\'{c} \inst{1,3}
          \and
          S. Vennerstrom \inst{4}
          \and
          G. Verbanac \inst{5}
          \and
          A. M. Veronig \inst{1,6}
          }

   \institute{University of Graz, Institute of Physics, Austria\\
              \email{stephan.heinemann@hmail.at}
         \and
            Centre for Plasma Astrophysics, K. U. Leuven, Belgium
         \and
             University of Zagreb, Faculty of Geodesy, Hvar Observatory, Croatia
         \and
             National Space Institute, DTU Space, Denmark
        \and
             University of Zagreb, Faculty of Science, Department of Geophysics, Croatia
         \and
             Kanzelh\"ohe Observatory for Solar and Environmental Research, University of Graz, Austria
             }

   \date{Received January 29, 2020; accepted April 21, 2020}

% \abstract{}{}{}{}{} 
% 5 {} token are mandatory
 
  \abstract
  % context heading (optional)
  % {} leave it empty if necessary  
   {The study of the evolution of coronal holes is especially important in the context of high--speed solar wind streams emanating from them. Slow and high speed stream interaction regions may deliver large amount of energy into the Earth magnetosphere-ionosphere system, cause geomagnetic storms, and shape interplanetary space.}
  % aims heading (mandatory)
   {By statistically investigating the long--term evolution of well observed coronal holes we aim to reveal processes that drive the observed changes in the  coronal hole parameters. By analyzing 16 long--living coronal holes observed by the \textit{Solar Dynamic Observatory}, we focus on coronal, morphological and underlying photospheric magnetic field characteristics as well as investigate the evolution of the associated high--speed streams.}
  % methods heading (mandatory)
   {We use the \textit{Collection of Analysis Tools for Coronal Holes} (CATCH) to extract and analyze coronal holes using $193$\AA\ EUV observations taken by the \textit{Atmospheric Imaging Assembly} as well as line--of--sight magnetograms observed by the \textit{Helioseismic and Magnetic Imager}. We derive changes in the coronal hole properties and correlate them to the coronal hole evolution. Further we analyze the properties of the high--speed stream signatures near 1au from OMNI data by manually extracting the peak bulk velocity of the solar wind plasma.}
  % results heading (mandatory)
   {We find that the area evolution of coronal holes mostly shows a rough trend of growing to a maximum followed by a decay. No correlation of the area evolution to the evolution of the signed magnetic flux and signed magnetic flux density enclosed in the projected coronal hole area was found. From this we conclude that the magnetic flux within the extracted coronal hole boundaries is not the main cause for its area evolution.  We derive coronal hole area change rates (growth and decay) of $(14.2 \pm 15.0) \times 10^{8}$ km$^{2}$ per day showing a reasonable anti--correlation (cc$_{Pearson}=-0.48$) to the solar activity, approximated by the sunspot number. The change rates of the signed mean magnetic flux density ($27.3 \pm 32.2$ mG day$^{-1}$)  and the signed magnetic flux ($30.3 \pm 31.5$ $10^{18}$ Mx day$^{-1}$) were also found to be dependent on solar activity (cc$_{Pearson}=0.50$  and cc$_{Pearson}=0.69$ respectively) rather than on the individual CH evolutions. Further we find that the coronal hole area--to-- high--speed stream peak velocity relation is valid for each coronal hole over its evolution but revealing significant variations in the slopes of the regression lines.}
  % conclusions heading (optional), leave it empty if necessary 
   {}

   \keywords{   sun --
                solar corona --
                EUV --
                observations --
                photosphere --
                magnetic fields
               }

   \maketitle
%
%-------------------------------------------------------------------

\section{Introduction}
Coronal holes (CH) are stable, long-lived structures usually observed in the solar corona as regions of reduced emission in extreme ultraviolet (EUV) or X-ray. CHs are characterized by an open magnetic field configuration along which ionized atoms and electrons are accelerated into interplanetary space, forming high speed solar wind streams \citep[HSS;][]{1973krieger,1976nolte,cranmer2009}. The interaction of HSSs with the preceeding ambient slow solar wind forms stream interaction regions (SIRs) which can develop into co-rotating interaction regions \citep[CIRs;][]{wilcox68,Tsurutani2006} if the CH persists over multiple rotations. These transient features are the main source of usually weak to medium geomagnetic storms at Earth \citep[see, \textit{e.g.},][]{alves06,verbanac11,2017vrshnak,2018yermolaev,2018richardson}. Since the Skylab-era, recurrent geomagnetic effects of CIRs have been associated with large scale coronal signatures, the CHs. \cite{1976nolte} first found an empirical relationship between the CH area and the peak bulk velocity of the solar wind. The linear CH area -- HSS speed relation is supported by many studies, however, different slopes of the regression lines are found \citep[\textit{e.g.,}][]{1978nolte,2007vrsnak,2017tokumaru,2018heinemann_paperI,2018hofmeister}.

The long-term evolution of recurrent CH structures and the associated HSSs has rarely been studied. Most studies focus on a snap-shot of the CH during its lifetime in form of a case study \citep[\textit{e.g.},][]{1999gibson,1999gopalswamy,2007Zhang,2009yang} or statistical studies that do not separate the evolution of individual CHs \citep[\textit{e.g.},][]{1982Harvey,2017Lowder,2017hofmeister,2019heinemann_catch}.  In the Skylab-era \cite{Bohlin1977} and \cite{Bohlin1978} proposed that the area and the mean magnetic flux density of a CH are dependent on its age and \cite{1978nolte,1978nolte_shorttermboundary} used X-ray observations to show that CHs primarily evolve by sudden, large-scale shifts in the position of the boundaries. A change in the underlying photospheric magnetic field is believed to influence the appearance of the CHs in the corona \citep{wang90,gosling96}. In a more recent study  \cite{Ikhsanov2015} showed that there are two magnetic field systems (large scale high latitude and small scale low latitude magnetic field systems) that play a main role in the evolution of the solar magnetic cycle and whose effects can be seen in the appearance of CHs. \cite{wang90} showed in their current--free coronal model, that the long-term evolution of CHs and the associated wind streams is related to flux emergence in active regions and the subsequent flux transport processes. CHs themselves seem to form in response to local (low-latitudes) or global (polar, high-latitudes) changes in the magnetic field topology. At low-latitudes, small scale magnetic systems may manifest themselves as localized CHs that are the result of decaying local sunspot groups \citep{2013petrie} or generated by other mechanisms that enable to \textit{open} the magnetic field, such as coronal mass ejections (CMEs)/filament eruptions, flux accumulation, or flux emergence \citep[\textit{e.g.},][]{webb78,2018heinemann_paperII}. \cite{2018heinemann_paperI,2018heinemann_paperII} give an in-depth view on the evolution of one particular CH in 2012 from which they identified a clear evolutionary pattern in combination with a correlation to the underlying magnetic field.

To statistically obtain the evolutionary aspects of individual CHs in the photosphere and corona the current study investigates in detail the evolution of 16 long-lived individual CHs. The observational CH data covers most of the \textit{Solar Dynamics Observatory} (SDO; \citealt{2012pesnell_SDO}) era so far, starting from September 2010 until April 2019. This enables us to obtain a more general insight into the evolutionary processes of CHs by analyzing coronal and magnetic properties as well as the rates at which they change. Additionally we investigate how changes in the observed CH on-disk characteristics affect the \textit{in--situ} measured HSS parameters near Earth as well as check if the well established CH area -- HSS peak velocity relation holds over individual CH evolutions. For this we used data from the Global Geospace Science Wind satellite \citep{1995acuna_GSS} and the \textit{Advanced Composition Explorer} (ACE; \citealt{1998stone_ACE}).

%--------------------------------------------------------------------
\section{Methods}\label{s:methods}

   \begin{figure*}[htbp]
   \centering
             \includegraphics[width=\textwidth]{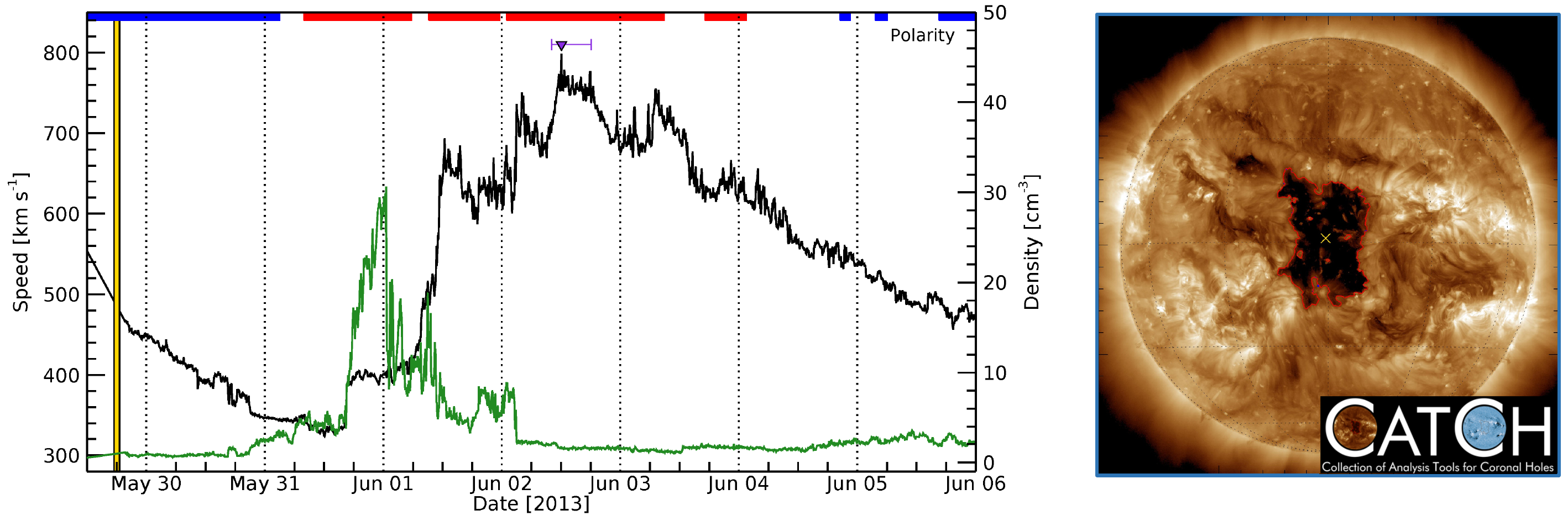}
       \caption{Example of an \textit{in--situ} signature of the solar wind data provided by the OMNI database (left), as a CH under study was observed near the center of the solar disk by SDO on May 29, 2013 at 18UT (right). The black line shows the solar wind bulk velocity and green line represents the plasma density. The purple triangle represents the peak velocity and the horizontal bar the average interval to show the plateau speed. The colored bar on the top represents the \textit{in--situ} polarity calculated after \cite{2002neugebauer}, with red being positive and blue negative polarity. The right panel shows a snapshot of the CH that is associated with this HSS, the time of the SDO observation corresponds to the yellow vertical line in the left panel. The red line is the extracted CH boundary (using CATCH) and the yellow x represents the center of mass of the CH. }
          \label{fig:insitu+ch_example}
          
   \end{figure*}
   
\subsection{Dataset}\label{subs:dataset}
For this study, we focus on the long-term evolution of CHs and its properties. In order to derive significant correlations or trends, we properly chose the dataset applying the following constraints: 

\begin{itemize}
    \item We aim not only to analyze intensity and morphological parameters of the CH evolution but also the change in the underlying photospheric magnetic field and enclosed magnetic flux. As such we are limited to observations for which remote sensing magnetic field data is available. For this reason, data from the \textit{Atmospheric Imaging Assembly} \citep[AIA;][]{2012lemen_AIA} and the \textit{Helioseismic and Magnetic Imager} \citep[HMI;][]{2012schou_HMI,2016couvidat_HMI} on board SDO was chosen for the analysis. SDO was preferred over the \textit{Solar and Heliospheric Observatory} \citep[SOHO;][]{1995soho} due to the significantly lower noise in the line--of--sight (LoS) magnetograms of HMI, which is very relevant for regions of low magnetic field strength such as CHs.
    \item To analyze the long--term evolution of CHs, only CHs which showed a clear coronal signature for at least five consecutive solar disk passages were considered.
    \item The last condition was, that the CH did not originate as or from a polar CH. This was considered due to high uncertainties in determining the CH area in polar regions due to projection and line--of--sight effects. Additionally the magnetic field observations of these areas are highly unreliable. 

\end{itemize}
By considering the stated constraints we extracted $16$ CHs fulfilling those requirements over most of the operational life-time of SDO from $2010$ to $2019$.

\subsection{Coronal Hole Detection and Extraction}\label{subs:ch-ex}
To identify and extract CHs we use the 193\AA\ filter of the AIA instrument on--board of SDO as CHs can be well observed in the emission of highly ionized iron (especially Fe \textsc{XII}: 193/195\AA). The high contrast to the surrounding quiet Sun enables a good extraction of boundaries. To capture the long--term evolution, we extract each CH and its characteristics at each solar disk passage near the central meridian. 
\cite{2018heinemann_paperI,2018heinemann_paperII} showed in a case study that one data point per rotation depicts the evolutionary trend sufficiently. We use the \textit{Collection of Analysis Tools for Coronal Holes} \citep[CATCH;][]{2019heinemann_catch}. CATCH uses a supervised threshold-based extraction that is modulated by the intensity gradient perpendicular to the CH boundary. Exemplary we show in the right panel of Figure~\ref{fig:insitu+ch_example} the extraction of a CH from May 29, 2013. In analogy to the right panel of Figure~\ref{fig:insitu+ch_example} we show in the Appendix (Figure~\ref{fig:appendix1}) for every CH a representative EUV image.

The CHs were extracted from point spread function (PSF) deconvoluted, limb-brightening corrected \citep{2014spoca} AIA 193\AA\ EUV images  that were rebinned to a pixel resolution $1024 \times 1024$, corresponding to a plate scale of $2.4$ arcseconds per pixel. Reducing the size increases processing speed and has only negligible effects on the boundary extraction \citep[\textit{e.g.}, see][]{2019heinemann_catch}. Furthermore, the boundaries have been smoothed using morphological operators with a kernel of $2\times2$ pixel size.

To account for LoS, projection and limb effects as well as for unreliable magnetograms near the poles we only consider pixels at latitudes $ | \lambda | < 60^{\mathrm{o}}$ in our calculations, \textit{e.g.}, all pixels above $60^{\mathrm{o}}$ latitude were disregarded.

The properties of the photospheric magnetic field underlying the extracted CH boundary were derived from de-projected 720s LoS HMI magnetograms which were chosen over its 45s equivalent due to the lower photon noise of $\approx 3$ Gauss near the disk center \citep{2016couvidat_HMI}. The higher signal--to--noise ratio gives more reliable results for low magnetic field densities. The extracted CH boundaries were overlaid on the magnetograms to calculate the properties.

\subsection{Coronal Hole Properties}\label{subs:ch-prop}

All properties were derived using CATCH, which includes morphological properties like the CH area, shape, position (center of mass) and properties of the underlying photospheric magnetic field like the signed mean magnetic flux density and magnetic flux. For more details about the extraction and calculation of the parameters and uncertainties we refer to \cite{2019heinemann_catch} and references therein.

In this study we focus on the evolution of the CH's primary parameters. The CH area (A), the signed mean magnetic flux density (B$_{s}$) and the signed magnetic flux ($\Phi_{s}$). The signed magnetic flux density is a good indicator for a CHs magnetic fine--structure as it is directly correlated to the abundance and coverage of strong unipolar magnetic elements within CHs. It has been found that they are the main contributors to the signed CH flux \citep{2019hofmeister,2019heinemann_catch}.

The average parameters over a CH's evolution, denoted by a "*", are derived by calculating the mean over the CH's evolution considering the uncertainties given by:
\begin{equation}
 P^{*}= \frac{1}{N} \sum^{N}_{i=0} P_{i},
 \end{equation}
 \begin{equation}
 \sigma^{*}=\sqrt{\frac{1}{N} \sum^{N}_{i=0} \sigma_{P_{i}}^{2} +\frac{1}{N-1} \sum^{N}_{i=0} (P_{i}-P*)^{2}}
\end{equation}
$P_{i}$ represents a CH property for every rotation ($i$) and $\sigma_{P_{i}}$ its uncertainty. The spread of the values over the whole evolution, $\sigma^{*}$, is calculated by adding the mean uncertainty (first term) and the standard deviation of the mean (second term). The change rates of CH properties were calculated as follows:
\begin{equation}
 dP_{i+1}=\frac{P_{i+1}-P_{i}}{t_{i-1}-t_{i}},   
\end{equation}
with P again representing a CH property (\textit{i.e.}, A, B$_s$, $\Phi_s$) and t being the time of observation.

\subsection{\textit{In--Situ} Data}\label{subs:insitu_methods}
To make more complete interpretation of the evolutionary effects of CHs on the \textit{in--situ} measured signatures, we investigated the associated HSSs near 1au using 5 minute plasma and magnetic field data provided by the OMNI\footnote{https://omniweb.gsfc.nasa.gov/} database. OMNI \textit{in--situ} measurements are propagated to the Earth’s Bow Shock Nose.

HSSs signatures were classified using the criteria given by \cite{jian06,jian09} and cross-checked with ready HSS lists (maintained by S. Vennerstroem for 2010--2019). The association of an identified HSS structure to an extracted CH was done (1) by comparing the dominant polarity of the HSS \citep[as given in][]{2002neugebauer} with the dominant polarity of the CH. Small scale turbulences (as reported in the Parker Solar Probe results by \citealt{2019bale}) were neglected. And (2) by considering the travel time of the HSS according to its speed \citep[\textit{e.g.},][]{2007vrsnak}. Time ranges covering ICME signatures were removed using ready-catalogs maintained by Richardson and Cane\footnote{http://www.srl.caltech.edu/ACE/ASC/DATA/level3/icmetable2.htm} for ACE \citep[see][for a description of the catalog]{2010richardson_RC-list}. CHs that could not be associated with a HSS signature were excluded from the HSS analysis. 

For each CH we derive, if possible, for each solar disk passage the associated HSS speed peak and the plasma density and magnetic field at the same time. From the peak we define the plateau speed as the averaged peak speed of the HSS within an interval of $[-2,+6]$ hours around the identified peak. The asymmetrical averaging interval was chosen to consider the asymmetry in the speed profile of HSSs and to avoid contribution from the stream interaction region. The plasma density and magnetic field magnitude for the identical time interval were also calculated in the same way. 

Here, the aim was to verify if and how the CH area -- HSS speed relation \citep{1976nolte,2007vrsnak,2012rotter,2017tokumaru,2018hofmeister} relates to the CH evolution. It has been found that there is a latitudinal dependence in CH area -- HSS speed relation and because we are interested in the CH area that affects Earth, the \textit{geoeffective CH area}, we correct for it. It can be calculated using the relation derived by \cite{2018hofmeister}:
\begin{equation}\label{eq:geoeffective_area}
    A_{\mathrm{geo}}=A_{CH} \times (1~-~|~\varphi_{co}(^{\mathrm{o}})~|~/~61.4),
\end{equation}
with $\varphi_{co}$ being the latitudinal separation angle between the center of mass of the CH and the observing spacecraft. According to this statistical relation, there is no contribution to the HSS from CHs above a latitude of $\approx 60^{\mathrm{o}}$, which seems to be valid in a first order approximation \citep{2007vrsnak}. Consequently, we consider CH areas below  $60^{\mathrm{o}}$.

\subsection{Correlations}\label{subs:corr_methods}
The analysis of different parameters, \textit{e.g.}, by correlation coefficients was done using a bootstrapping method \citep{efron1979_bootstrap,efron93_bootstrap} to derive errors that take into account the low sample size within each evolution (between $5$ and $18$).

\section{Results} \label{s:results}
By examining a sample of $16$ long--lived CHs throughout the SDO-era and its associated HSSs we obtained the following results:

\subsection{Lifetime}\label{subs:lifetime}
\begin{figure*}[htbp]
   \centering
             \includegraphics[width=\textwidth]{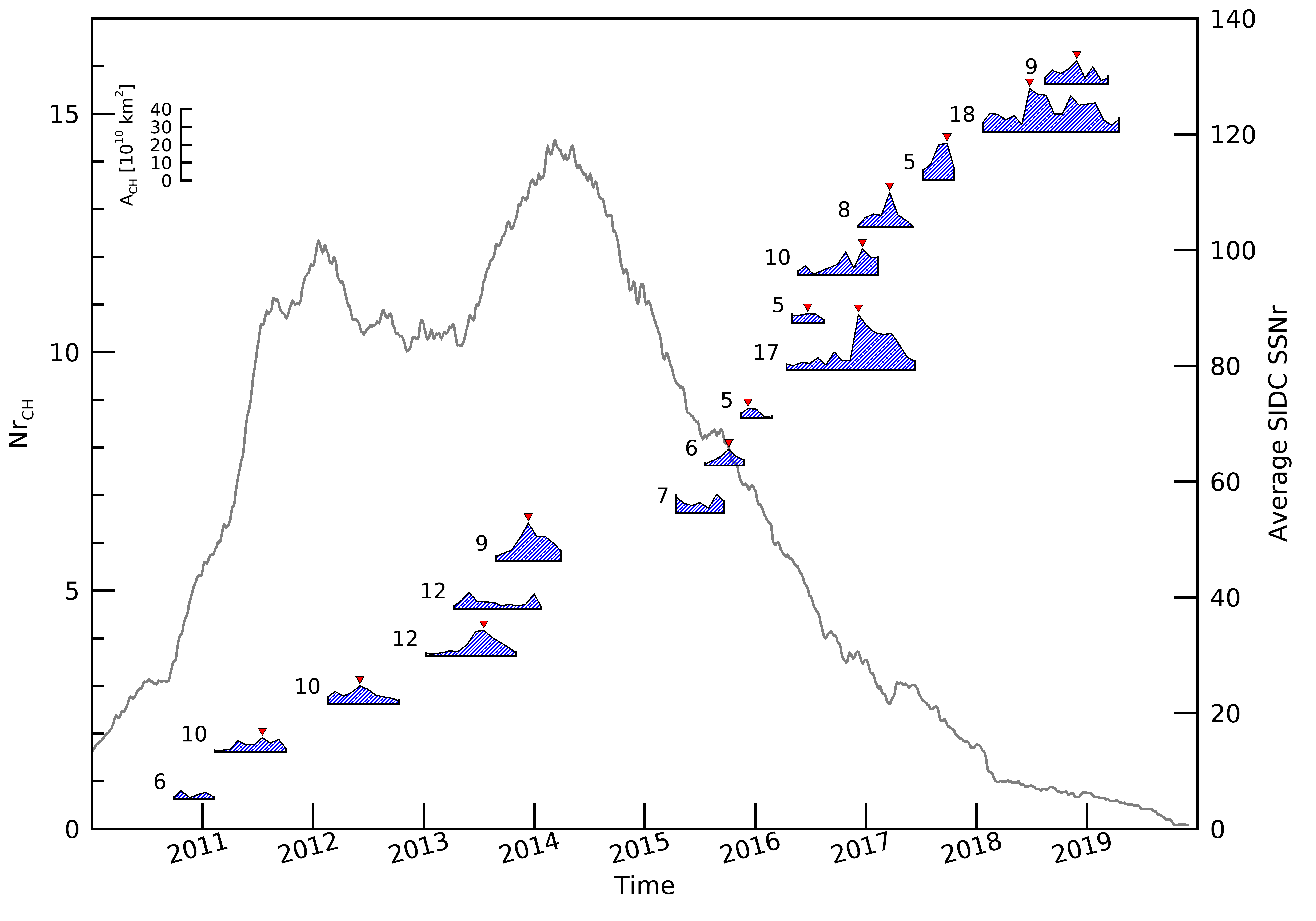}
       \caption{Temporal evolution of all the CHs under study. The length of the bars represents the lifetime from the first to the last observation near the central meridian. The profile displays the area evolution (scale in the left corner of the figure). The red triangle marks the peak in the CH area. If no clear central peak was detected, the mark was omitted. The number in front of the bars shows how many times the CH was observed passing the central meridian. In the background, the solar activity, approximated by the smoothed sunspot number (Source: WDC-SILSO, Royal Observatory of Belgium, Brussels), can be seen.}
          \label{fig:ch_overview}

   \end{figure*}

We define a CH's life from the time of its first observed central meridian passage to its last. This gives a lower boundary of the lifetime of the analyzed CHs. The accurate time of birth and death of a CH is generally a rarely observed phenomenon, mostly due to the limited coverage of the solar surface. Figure~\ref{fig:ch_overview} depicts the 16 recurrent CHs that were under study. The length of the bar represents the lifetime, the position the date, and the number gives the count of rotations the CH was observable. The profile depicts the individual CH area evolution. In the background the smoothed sunspot number provided by WDC-SILSO\footnote{Royal Observatory of Belgium, Brussels: http://www.sidc.be/silso/}, a proxy for the solar activity, is shown. 

CHs having a lifetime from $5$ up to $18$ rotations were observed. The 2 CHs with the significantly longest lifetimes ($17$ and $18$ rotations) were observed during the solar minimum ($2017-2019$) but this is obviously not a general trend. Note, that this only regards CHs already considered long--living and in a sample where polar CHs are excluded. We nearly continuously observe long--living CHs throughout the solar cycle with only one notable exception near the maximum in the solar activity (between $2014$ and $2015$).

\subsection{Coronal Hole Evolution in A, B$_{s}$ and $\Phi_{s}$}\label{subs:evo}
   \begin{figure*}[htbp]
   \centering
             \includegraphics[width=0.85\textwidth]{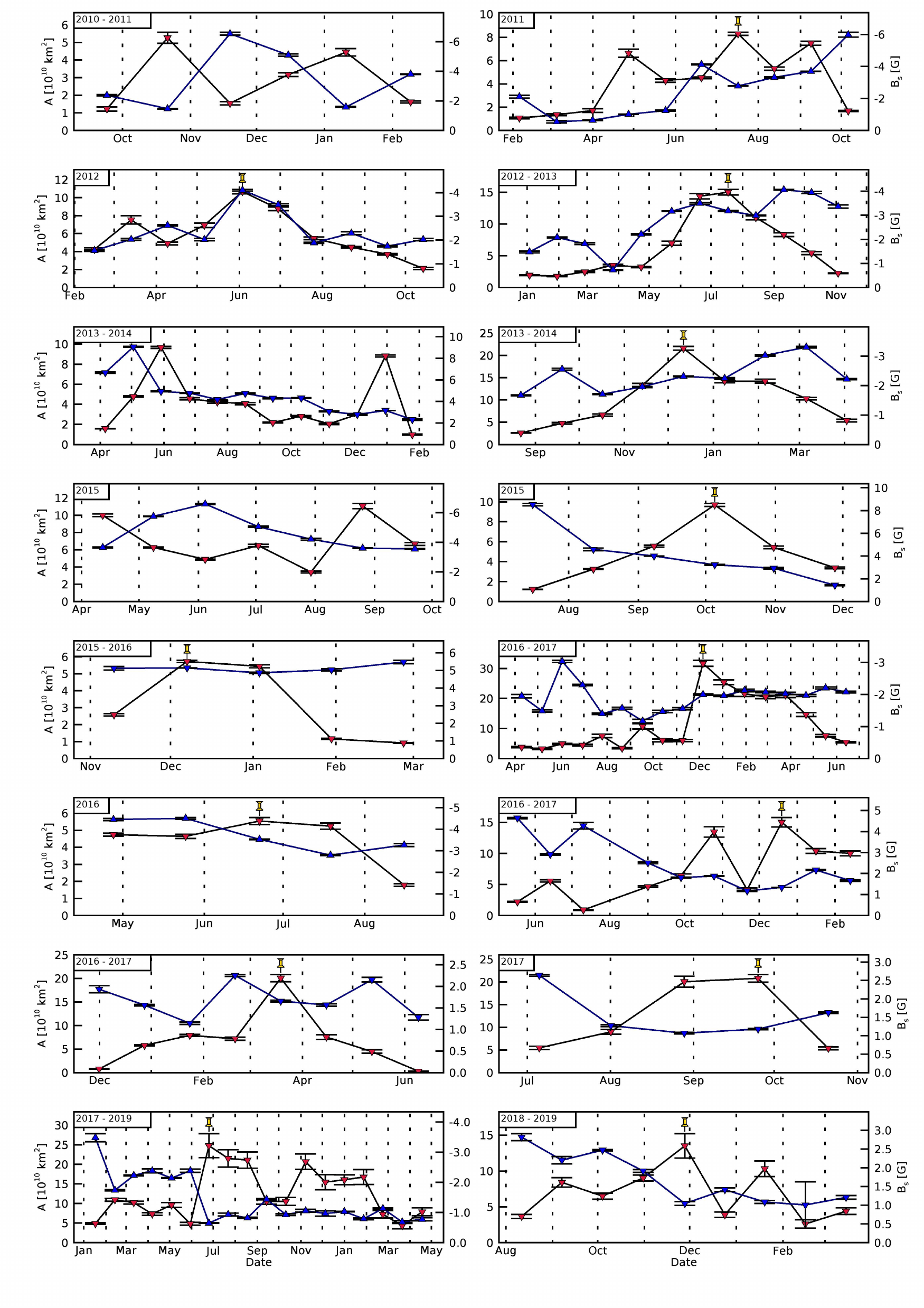}
       \caption{ Evolution of the area (black--red) and magnetic flux density (blue) of all CHs under study. The error bars represent the uncertainties. The vertical dashed guidelines represent the first day of each month and the yellow pin marks the peak in the observed CH area (if a clear peak was associated).}
          \label{fig:full_area}
          
   \end{figure*}

   \begin{figure}[htbp]
   \centering
             \includegraphics[width=9cm]{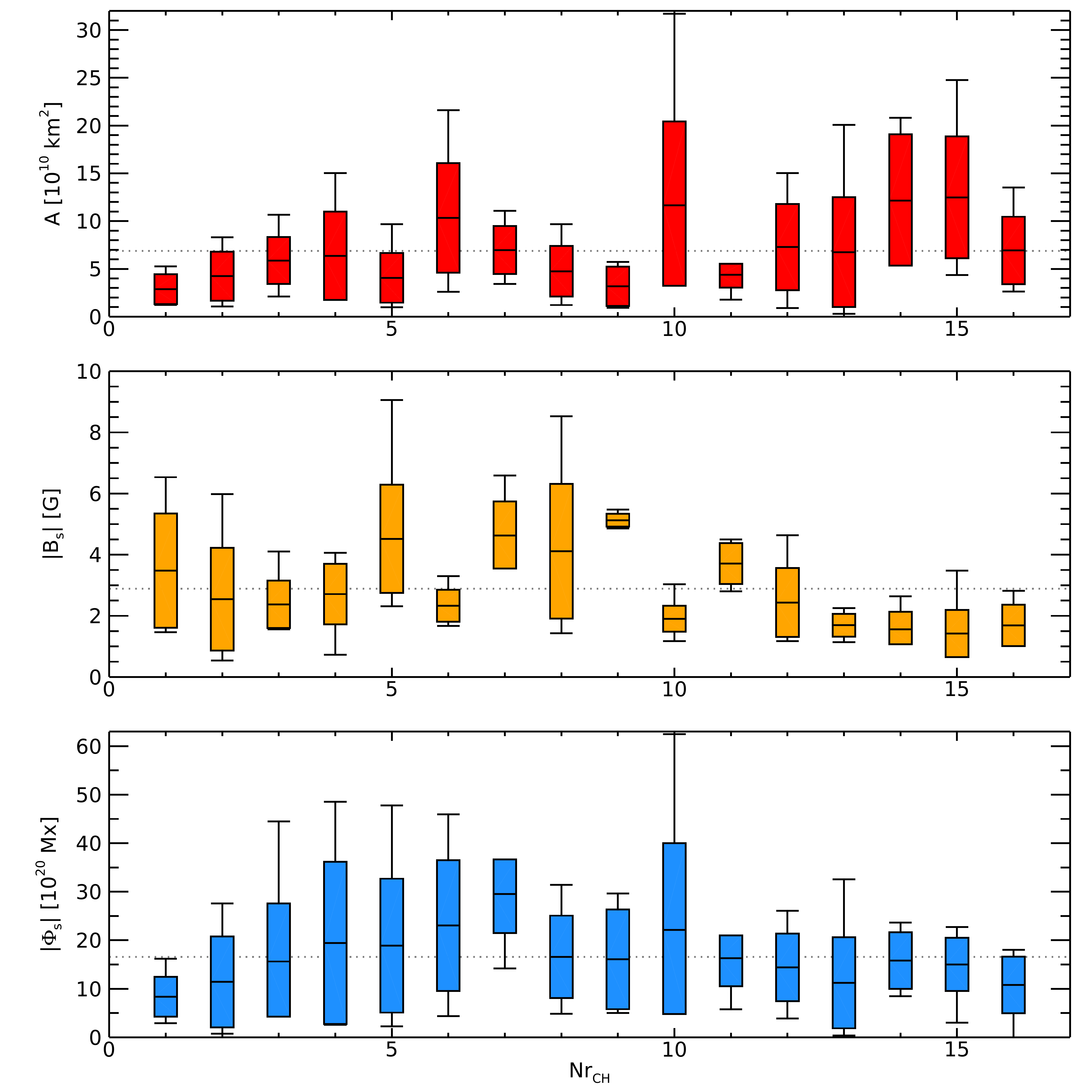}
       \caption{Average properties over each CHs evolution. The middle represents the average property (A*,B$_{s}$*,$\Phi_{s}$*), the colored bar the $1\sigma$ range and the whiskers show the full range of values the parameter observed over the evolution. The top panels shows the CH area, the middle the CH flux density and the bottom panel the signed flux.}
          \label{fig:prop}
          
   \end{figure}

By examining a sample of $16$ long--lived CHs throughout the SDO-era, we found that $13$ show a steady change in area, characterized by an approximately monotonic increase in the area followed by a roughly monotonic decrease (with scatter). Such a general trend is often visible, but the relative position of the peak, the maximum and minimum area, and the lifetime significantly vary. This shows the individuality in each CH's evolution. Figure~\ref{fig:full_area} shows the evolution of the CH area (black--red) of all 16 CHs studied. The blue line represents the evolution of the photospheric magnetic flux density within the projected CH area. Note, that large jumps in the area might be caused by merging/splitting with/into CHs, connection to polar CHs or nearby filament eruptions that abruptly open additional field lines.

CHs are usually considered to be large scale structures that are defined by their magnetic field topology, the predominantly open field configuration. Thus, we analyze the photospheric magnetic field underlying the extracted boundary to investigate the relation of magnetic field evolution to the area evolution. As shown in Figure~\ref{fig:full_area} the evolution of the mean signed magnetic flux density of different CHs does not follow a uniform trend but displays various kinds of behaviour. We find CHs that seem to have a visual correlation of area and flux density evolution as well as some with a supposed anti--correlation and ones without noticeable correlation. We quantify this later.

Figure~\ref{fig:prop} shows the average properties over each CHs evolution. On the top the CH area, in the middle the flux density and on the bottom the signed flux. We find only mild variations in the mean parameters (A*,B$_{s}$*,$\Phi_{s}$*) represented by the mid--point of each vertical bar. The varations over each evolution however may strongly vary. The CH areas over the entire CH evolution may vary slightly ($\sim 2-3 \times 10^{10}$ km$^{2}$) or strongly ($> 10 \times 10^{10}$ km$^{2}$) around the mean. We find magnetic flux densities ranging from $\sim 1$ G up to $\sim 9$ G with usual variations over a CHs lifetime of about $\pm (1-2)$ G around the mean. The value of the mean magnetic flux of CHs over their evolutions seems to be similiar for all CHs, however the variation throughout the evolution seem to be in the order of the mean.

\subsection{Change Rates}\label{subs:mag-evo}
   
     \begin{figure}[htbp]
   \centering
             \includegraphics[width=9cm]{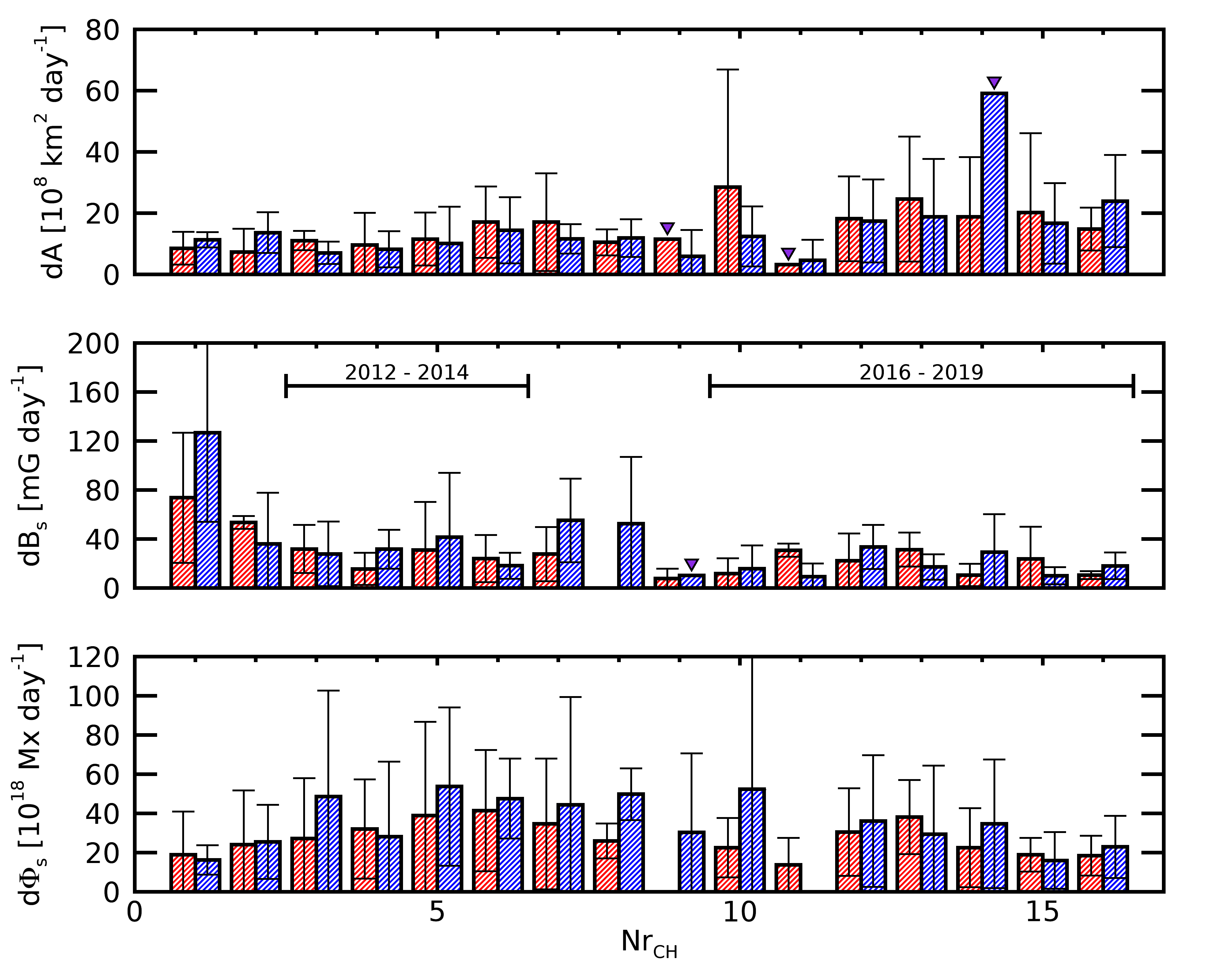}
       \caption{Average change rates (growth: red; decay: blue) for each CH evolution individually. The top panels hows the average CH area change rates, the middle the CH flux density change rates  and the bottom panel the change rates of the  signed flux. The purple triangles mark values which have been calculated from only 1 or 2 measurement points and are as such more unreliable. }
          \label{fig:area_growth}
          
   \end{figure}

Apart from the evolutionary profile in the CH properties we investigated the average rates at which CH properties change (a) to evaluate if CHs evolve at similar rates or differ significantly from each other and (b) as an important information on how the open magnetic field changes as CHs are a major contributor of open magnetic flux (in regards to the open flux problem). Figure~\ref{fig:area_growth} shows the average change rates of the individual CH evolutions separated into growth rates (positive change rates, red) and decay rates (negative change rates, blue). Averaged over all CH evolutions, we find an area change rate (Figure~\ref{fig:area_growth}, top panel) of $(14.2 \pm 15.0) \times 10^{8}$ km$^{2}$ day$^{-1}$ that can be divided into an area growth rate of $ (15.4 \pm 18.5) \times 10^{8}$ km$^{2}$ day$^{-1}$ and a similar area decay rate of $(13.2 \pm 12.6) \times 10^{8}$ km$^{2}$ day$^{-1}$. This suggests that, on average, the CH grows and decays at a similar rate. We observe a possible solar cycle dependence of the area change rates as in and near the solar maximum ($2012-2014$) we find change rates of $(10.7 \pm 8.9) \times 10^{8}$ km$^{2}$ day$^{-1}$. During the descending phase and solar minimum ($2016-2019$) the average area change rates are higher at around $(17.5 \pm 18.7) \times 10^{8}$ km$^{2}$ day$^{-1}$.

The second panel in Figure~\ref{fig:area_growth} shows the change rates for the mean signed magnetic flux density. We find similar rates for both growth and decay in each CHs evolution, on average $27.3 \pm 32.2$ mG day$^{-1}$. We again observe a difference in the rates between CHs near solar maximum (2012-2014) and CHs near the solar minimum (2016-2019). During the former, the flux density growths at an average rate of $ 24.5 \pm 25.1$ mG day$^{-1}$ and decays at a similar rate of $ 30.9 \pm 34.5$ mG day$^{-1}$. During the later interval the change rates drop by up to factor 2 to $ 17.8 \pm 18.8$ mG day$^{-1}$ for growth and $17.6 \pm 16.8$ mG day$^{-1}$  for decay.  

From the bottom panel of Figure~\ref{fig:area_growth} we can deduce that a CHs magnetic flux,  on average, evolves at a rate of $(30.3 \pm 31.5) \times 10^{18}$ Mx day$^{-1}$. Interestingly we find that for most CHs the average negative flux change rate is between $20-30\%$ higher than the average positive change rate.  It is not clear if this an effect dominated by a CHs area evolution or magnetic field evolution.

All change rates of CH area, mean signed magnetic flux density and magnetic flux are listed in Table~\ref{tab:change-rates}.

 \begin{table}
 \caption{Average change rates throuout the CH evolution. "max" denotes the time intervall from $2012-2014$ and "min" the interval $2016-2019$.}             % title of Table
 \label{tab:change-rates}      % is used to refer this table in the text
 \centering                          % used for centering table
 \begin{tabular}{l | c | c | c }        % centered columns (4 columns)
 \hline\hline                 % inserts double horizontal lines
 & dA & dB$_{s}$ & d$\Phi_{s}$ \\% table heading 
 & $10^{8}$ km$^{2}$ day$^{-1}$ & mG day$^{-1}$ & $10^{18}$ Mx day$^{-1}$ \\% table heading 
 \hline                        % inserts single horizontal line
  Total       & $14.2 \pm 15.0$ & $27.3 \pm 32.2$ & $30.3 \pm 31.5$\\      % inserting body 
  Growth         & $15.4 \pm 18.5$ & $24.4 \pm 27.1$ & $26.8 \pm 25.9$\\      % inserting body
  Decay          & $13.2 \pm 12.6$ & $29.6 \pm 37.6$ & $34.0 \pm 40.7$\\      % inserting body
   \hline   
  Total$_{max}$  & $10.7 \pm 8.9$  & $28.1 \pm 29.6$ & $38.4 \pm 36.1$\\      % inserting body
  Growth$_{max}$    & $12.2 \pm 9.8$  & $24.5 \pm 25.1$ & $34.4 \pm 35.9$\\      % inserting body
  Decay$_{max}$     & $9.5 \pm 8.9$   & $30.9 \pm 34.5$ & $43.0 \pm 40.4$\\      % inserting body
   \hline   
  Total$_{min}$  & $17.5 \pm 18.7$ & $17.7 \pm 17.0$ & $26.4 \pm 29.9$\\      % inserting body
  Growth$_{min}$    & $19.8 \pm 23.6$ & $17.8 \pm 18.8$ & $22.6 \pm 16.5$\\      % inserting body
  Decay$_{min}$    & $15.6 \pm 15.2$ & $17.6 \pm 16.8$ & $30.1 \pm 43.8$\\      % inserting body
 \hline \hline                                     %inserts single line
 \end{tabular}
 \end{table}

\subsection{Correlations}\label{subs:corr}

      \begin{figure}[htbp]
   \centering
             \includegraphics[width=9cm]{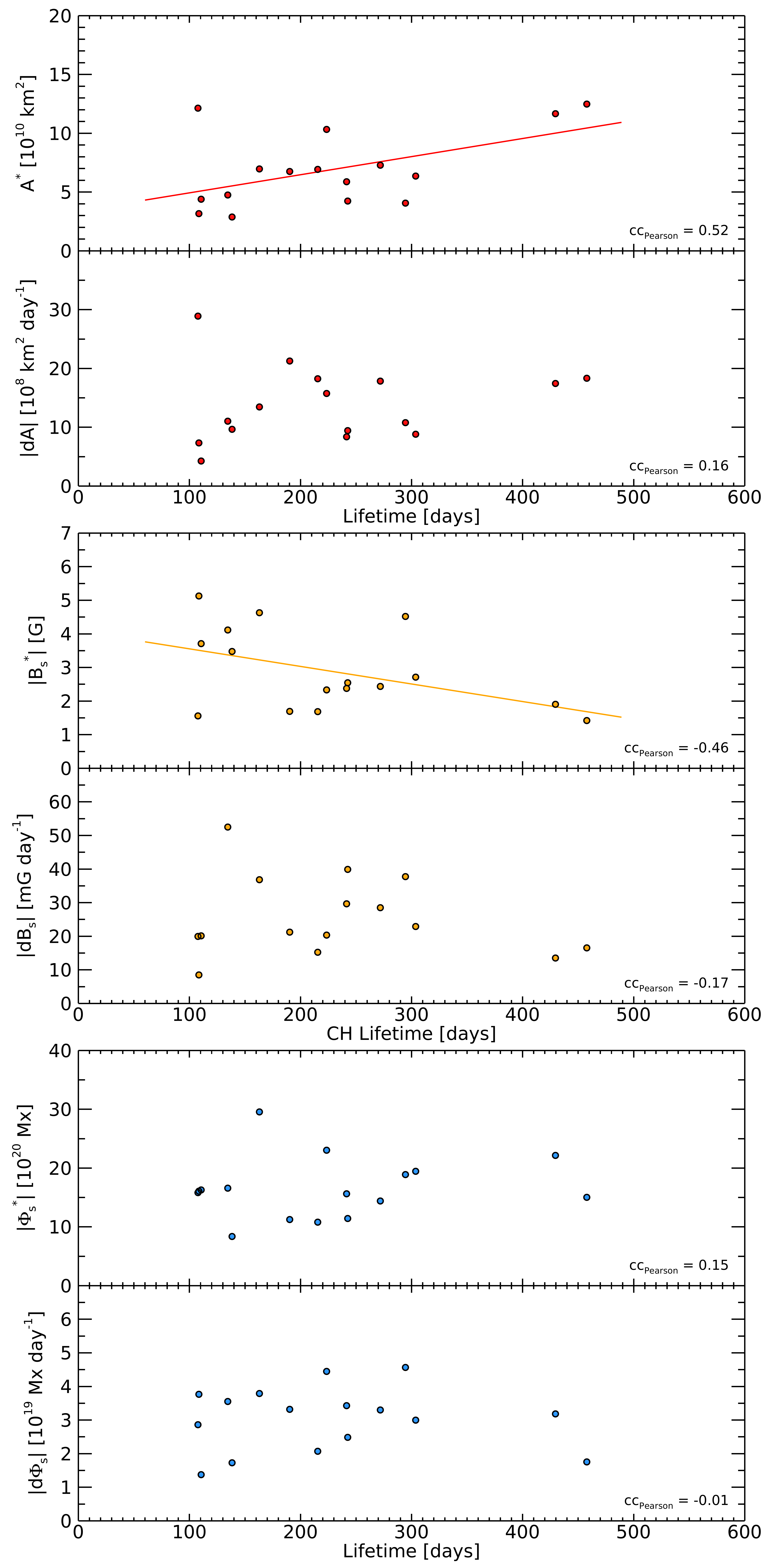}
       \caption{Scatterplot of the average evolutionary CH properties and their absolute change rates against the CH lifetime. The top panel (red) shows the CH area (A*, |dA|), middle panel (orange) displays the mean signed magnetic flux density (B$_{s}$*, |dB$_{s}$|) and the bottom panel (blue) represents the signed magnetic flux ($\Phi_{s}$*, |$d\Phi_{s}$|).}
          \label{fig:corr_lifetime}
          
   \end{figure}
         \begin{figure}[htbp]
   \centering
             \includegraphics[width=9cm]{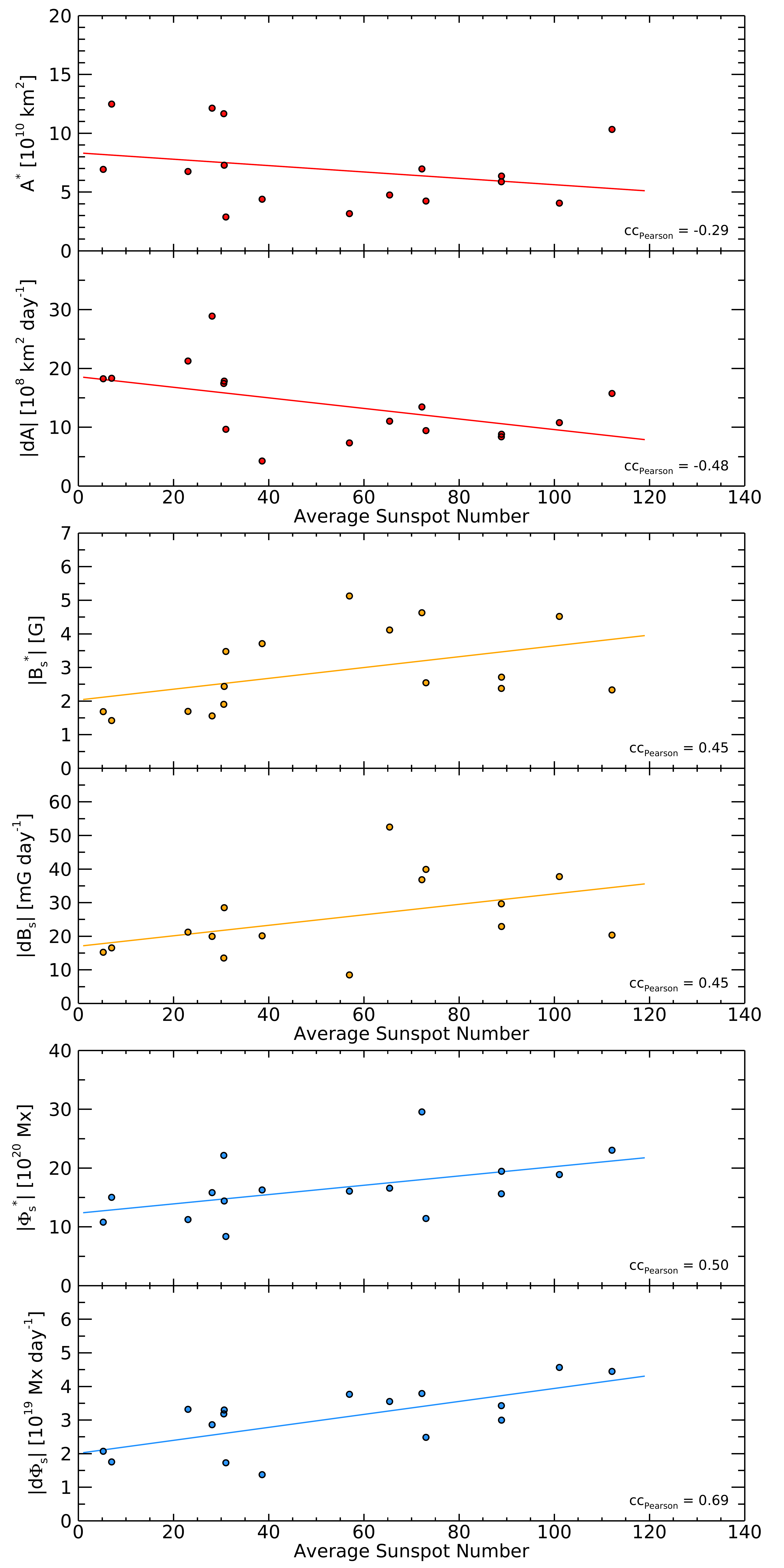}
       \caption{Scatterplot of the average evolutionary CH properties and their absolute change rates against the average sunspot number as a proxy for solar activity. The top panel (red) shows the CH area (A*, |dA|), middle panel (orange) displays the mean signed magnetic flux density (B$_{s}$*, |dB$_{s}$|) and the bottom panel (blue) represents the signed magnetic flux ($\Phi_{s}$*, |$d\Phi_{s}$|).}
          \label{fig:corr_ssnr}
          
   \end{figure}

   \begin{figure}[htbp]
   \centering
             \includegraphics[width=9cm]{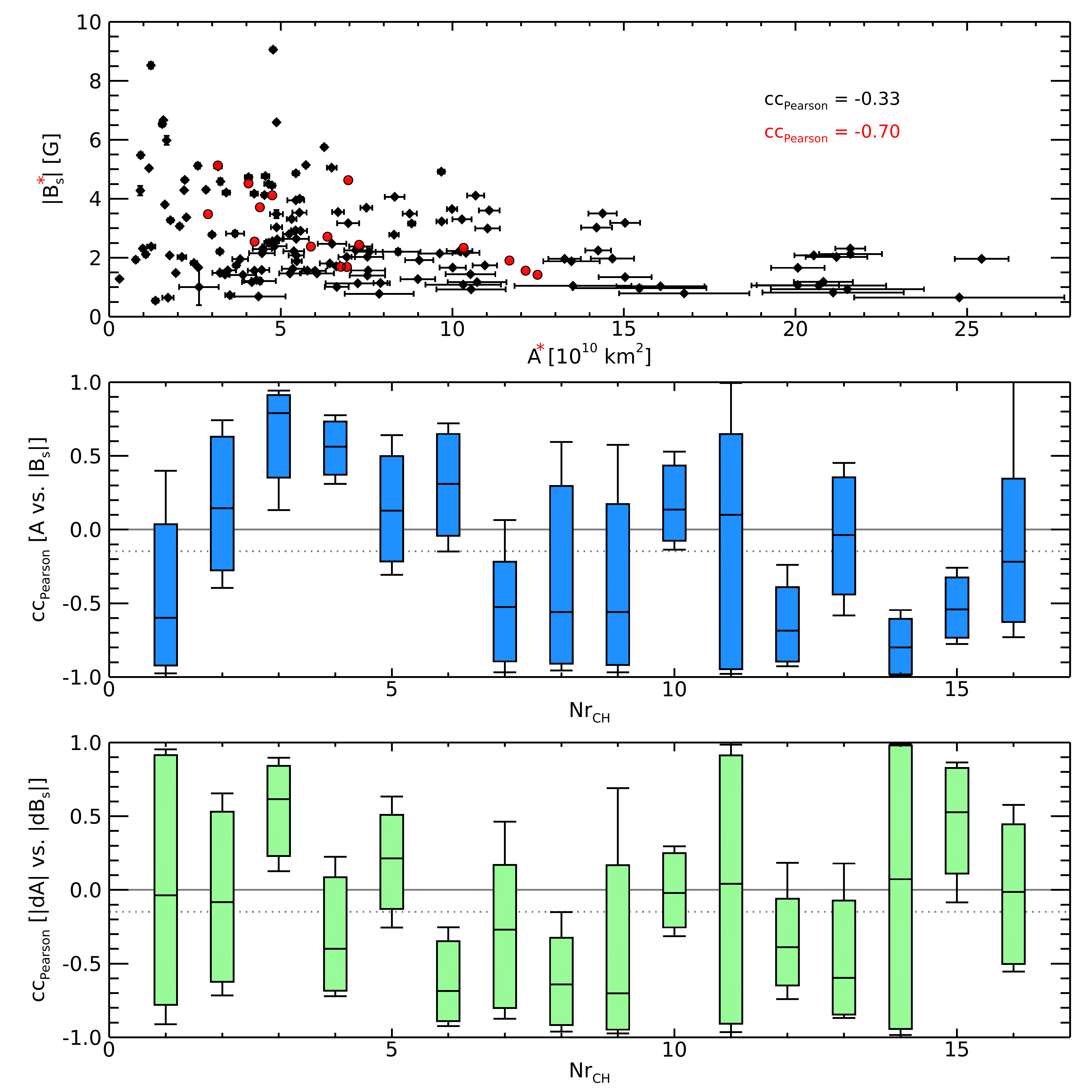}
       \caption{ Scatterplot of the absolute value of the signed mean magnetic flux density of CHs as function of the CH area. The top panel represents all observations (in red the average values over each evolution), the mid panel each CHs evolution individually and the bottom panel shows the correlation of the change in the area to a change in the magnetic flux density for each CH evolution individually as well. In the middle and the bottom panel the median of the bootstrapped sample for each evolution is represented by the horizontal line in the bar, the green and blue bars represent the $80\%$ percentiles and the whiskers the $90\%$ percentiles. The dotted line gives the mean of the median values.  }
          \label{fig:mag}
          
   \end{figure}

After deriving various evolutionary CH parameters we investigated possible correlations. Figure~\ref{fig:corr_lifetime} shows the three primary CH properties (averaged over the CH lifetime), namely area (top, red), flux density (middle, orange) and flux (bottom, blue), and their change rates as function of CH lifetime. We find moderate correlations for the average CH area ($cc_{Pearson}=0.52$) and the average flux density ($cc_{Pearson}=-0.46$), however strongly influenced by the two very large and long living CHs during solar minimum (2 datapoints on the far right). No correlation with the CH lifetime is visible in the change rates nor in the average magnetic flux. 

Figure~\ref{fig:corr_ssnr} investigates a possible dependence of average CH parameters and change rates on solar activity. As a proxy for the solar activity, the average sunspot number was used. We define the average sunspot number of a CHs evolution as the mean of SIDC sunspot number over the time interval in which the CH was observed. We find weak (anti-)correlations to solar activity of the average CH area and flux density as well as their change rates. Parameters that show the highest correlation with the solar activity are the average magnetic flux ($cc_{Pearson}=0.50$) and the flux change rate ($cc_{Pearson}=0.69$).

In Figure~\ref{fig:mag} we investigated a possible correlation of the CH area to the flux density. The top panel shows the flux density as function of the CH area for all individual CHs (black) and the average over each evolution (red). We find a weak anti--correlation for the whole sample ($cc_{Pearson}=-0.33$) and a strong anti--correlation for the averages ($cc_{Pearson}=-0.70$). Note, that the strong anti--correlation of the average CH area and mean magnetic flux density is due a smoothing effect of the whole sample and not a causal relation. The middle panel shows the correlation coefficients for each individual CH evolution. We find CHs in which the evolutionary profile of the area seems quite strongly correlated with the flux density evolution ($3/16$), some that show no correlation at all ($6/16$) and about half that show a weak to strong anti--correlation ($7/16$). We note here that the slopes of the regression line of some CH evolutions are very low due to an only slightly changing magnetic flux density throughout the lifetime. This can artificially enhance the Pearson correlation coefficient without a causal relation being present. All CHs retain the same polarity over their evolution. The change rates of CH area and flux density (Figure~\ref{fig:mag},bottom panel) do not seem to be correlated (average over all evolutions: $cc_{mean}=-0.15$).

\subsection{High Speed Stream Velocity Evolution}\label{subs:hss-evo}
   \begin{figure}[htbp]
   \centering
             \includegraphics[width=9cm]{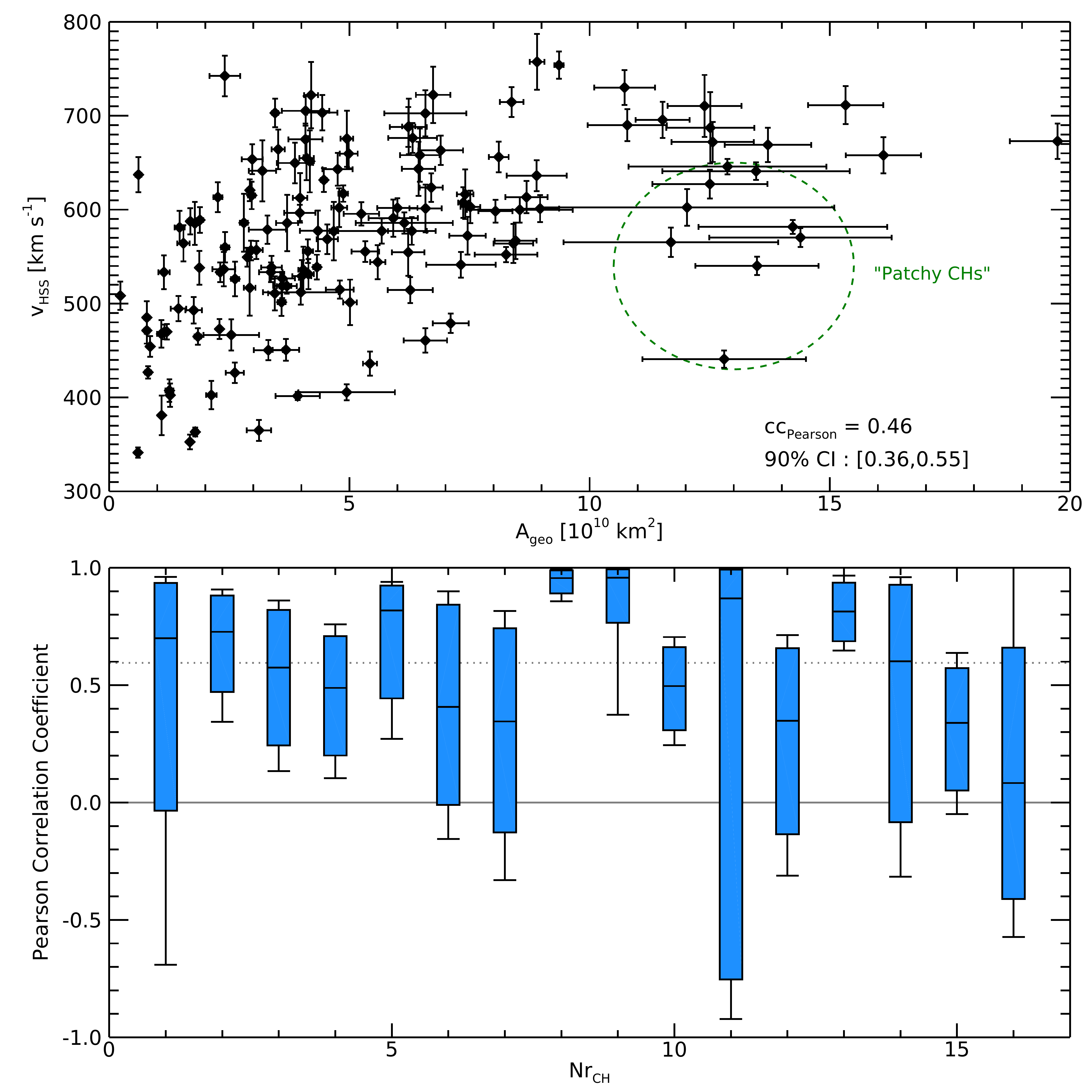}
       \caption{ Correlation plot of the CH area vs the plateau bulk velocity of the associated HSS. The top panel represents all observations and the bottom panel each CHs evolution individually. The green circle marks the "patchy" CHs which have large uncertainties in their area. In the bottom panel the median of the bootstrapped sample for each evolution is represented by the horizontal line in the bar, the green and blue bars represent the $80\%$ percentiles and the whiskers the $90\%$ percentiles. The dotted line gives the mean values. }
          \label{fig:hss}
          
   \end{figure}

 Here we investigate if the well--established CH area -- HSS peak velocity relation is visible over individual CH evolutions. Using the \textit{geoeffective CH area}, calculated from Equation~\ref{eq:geoeffective_area}, we derive a Pearson correlation coefficient of $cc_{Pearson}=0.46$ with a $90\%$ confidence interval of $[0.36,0.55]$ for the full sample. If considering each evolution individually we find a large range of correlation coefficients between medium and strong correlations (see Fig.~\ref{fig:hss}). Possible reasons for this will be discussed later. The mean correlation coefficient of all 16 individual evolution correlation coefficients comes to  $cc_{mean} =  0.59 \pm 0.26 $. The slopes of the individual regression lines vary largely around $(3 -44)\times 10^{-10}$ km$^{-1}$ s$^{-1}$ with accordingly different y--intercepts. For large CHs (A $> 10 \times 10^{10}$ km$^{2}$) we find a saturation effect in the peak velocity. For "patchy" CHs (large fragmented CHs with weakly defined boundaries) the maximum peak velocity seems to be lower than one would expect for CHs of such area.
We investigated \textit{in--situ} measured plasma density and magnetic field magnitude at the position of the velocity plateau and did not find any correlations to the CH area or magnetic field properties.

\section{Discussion} \label{s:discussion}
Using EUV full--disk observations by AIA and HMI magnetograms throughout most of the operational lifetime of SDO between $2010$ and $2019$ we studied the evolution of a sample of $16$ long--living CHs and their associated HSSs.

We find that $13$ out of $16$ CHs follow a visible trend in the area evolution covering roughly a growing phase until a clear maximum is reached followed by a decay phase. Only $3$ CHs show an erratic behavior, \textit{e.g.,} revealing multiple peaks in the area or a maximum area at the birth or decay. From this we deduce that the majority of CHs ($>80\%$) reveal an area evolution that follows a pattern of growth and decay. However, we note that due to the small sample we cannot comment on the significance. The proposed 3-phase evolution \citep{2018heinemann_paperI} might be rather the exception as it was only clearly observed for one CH.

We find a moderate correlation of the average CH area and flux density but the correlation is not significant. Also, the change rates seem independent of the lifetime.

We find a correlation of the average CH properties as well as change rates with the solar activity. Especially the change rate of the magnetic flux in CHs is strongly correlated to the average sunspot number during the CH life time. We suspect that depending on magnetic activity (magnetic pressure) in quiet sun areas and active region, near the CH, the open field is able to expand accordingly. 

Specifically, we found that near and during solar maximum ($2012-2014$), the area change rate was found to be $(10.7 \pm 8.9) \times 10^{8}$ km$^{2}$ day$^{-1}$ (similar for both growth and decay rates) and during solar minimum the rate was around $60\%$ larger at  $(17.5 \pm 18.7) \times 10^{8}$ km$^{2}$ day$^{-1}$. Our results for the solar maximum match the area change rates found by \cite{Bohlin1977} with $(13.0 \pm 3.5)~10^{8}$ km$^{2}$ day$^{-1}$ quite well. \cite{1978nolte} used Skylab X--ray observations  of long--living coronal holes during the descending phase of solar cycle 20 and derived rates of $(7.3 \pm 1.4)~10^{8}$ km$^{2}$ day$^{-1}$, which is lower than the values found in our study. Large deviations between the rates from previous studies can be primarily found in the CH evolutions during low solar activity where much larger average rates were derived. The growth and decay rates were found to be rather stable between long--living CHs observed during times of similar solar activity. The consistent average rates suggest that discrete events that may cause sudden large--scale changes in the CHs, like filament eruptions that permanently open magnetic field lines \cite[\textit{e.g.,}][]{1990Kahler}, are rare events and that CHs evolve more gradually. This might explain the discrepancy between \cite{1978nolte}, who claimed that most of a CHs area evolution can be explained by discrete boundary changes of sizes larger than $2.7~ 10^{8}$ km$^{2}$, and \cite{1990Kahler} who found no such sudden large--scale changes. When comparing the average rate at which CHs evolve with results from flux transport calculations by \cite{1964Leighton} we are in good agreement during the solar maximum and off during low solar activity.

A similar picture can be deduced from the change rates of the mean signed magnetic flux density enclosed within the projected CH boundary. For evolutions of individual CHs the rate at which the flux density changes stays at rather similar values but the percentual scatter is larger than in the area change rates. We show that the change rates of the flux density is higher during enhanced solar activity, for the timerange of $2012-2014$ (near and at solar maximum) we derive rates of ($ 28 \pm 30$ mG day$^{-1}$) and over $30\%$ lower rates of $ 18 \pm 17$ mG day$^{-1}$ near solar minimum ($2016-2019$). This hints towards a small imbalance between flux appearance and flux removal rates as well as low rates in general. \cite{2016gosic} found average net flux density change rates of $5.0 \pm 5.8$ G day$^{-1}$ in internetwork elements of two supergranular cells in the quiet sun using $38$ hour continuous HINODE observations. \cite{2017Smitha} report values about a factor 10 larger using SUNRISE data. We note, that comparison should be treated cautiously as the rates derived by both \cite{2016gosic} and \cite{2017Smitha} are from internetwork elements and the ones in this study from the total extracted CH area. Also, HINODE and SUNRISE obtain their magnetograms in a much higher resolution and sensitivity than HMI/SDO which might explain the difference to the results gathered in this study.

\cite{2007Zhang} found that in a decaying CH a decrease of flux is caused by the emergence of opposite polarity flux within the CH boundaries and also that nearly no flux transport across the CH boundary is observed. According to this, changes in the magnetic flux density in CHs are caused by processes within the CH boundaries, \textit{e.g,}  flux emergence rate. The average CH area and average flux density over the CH evolutions seem to be correlated, however this is simply a smoothing effect and does not imply a causal correlation (see Figure~\ref{fig:mag}). From the statistical distribution of small unipolar magnetic elements within CHs, that have been found to define a CHs magnetic topology, it can be simply deduced that only smaller CHs might have higher flux densities (but also can have small flux density). Whereas in large CHs the flux density tends to converge to a mean flux density of about $1-3$ G \citep[see ][]{2018heinemann_paperII,2019heinemann_catch,2019hofmeister}. As such the derived anti--correlation is no evolutionary effect. Larger statistical studies of CH properties show no correlation between the CH area and its magnetic flux density \citep{2017hofmeister,2019heinemann_catch}.

As such the evolution of the magnetic flux density and magnetic flux cannot explain the evolutionary behaviour of the CH area. We propose that the area evolution of CHs is not primarily caused by the evolution of the mean signed magnetic flux density within the projected CH boundary as extracted from EUV observations. Possible processes that might play a role in the area evolution include interchange reconnection \citep[see][]{1984Shelke,2004wang,2004Madjarska,2005Fisk,2009madjarska,2010edmondson,2011krista,2011yang,2014ma,2018kong} and/or the change in the global and local magnetic field configuration outside the CH.

Additionally to the evolutionary trend derived from remote sensing data, we investigated the evolution of the \textit{in--situ} measured peak velocity of the associated HSS. Using the \textit{geoeffective CH area}, we derive a mean correlation coefficient of the individual CH evolutions of  $cc_{Pearson,mean} =  0.59 \pm 0.26 $, which is in the range of previous statistical studies of the CH area -- HSS peak velocity relation (\citealt{2011Karachik}: $cc_{Pearson}= 0.41$ to $0.65$; \citealt{2007vrsnak,2009Abramenko,2011verbanac_b,2011verbanac_a,2012rotter,2017tokumaru,2018hofmeister}): $cc_{Pearson}= 0.62$ to $0.80$). We are also in good agreement with \cite{2018heinemann_paperI}, who found in their case study of a long--living CH a correlation coefficient of $cc_{Pearson}= 0.77$. We find a saturation effect in the peak velocity -- CH area relation, where the peak velocity of large CHs does not follow a linear trend but rather converges to a speed of roughly $700$ km s$^{-1}$. This might be related to the longitudinal saturation found by \cite{2018garton}.

We can conclude that the CH area -- HSS peak velocity relation is clearly visible within each CH's evolution, but the slopes of the linear regression lines vary strongly. We identify 3 possible causes for the varying slopes: (1) The \textit{in--situ} measured bulk velocity depends not only on the CH but also on the preconditioning of the interplanetary space \citep[\textit{e.g.},][]{2017temmer-precond}. HSS or CMEs near or before the arrival of the HSS under study can significantly influence the measured speeds. This may lead to an enhanced scatter in the relations. (2) The relation between CH area and HSS peak velocity is individual for each CH. (3) It might be possible that the CH area and the HSS peak velocity are not directly causally related.. It is known that geometry of CHs plays an important role in in defining the 3D shape of the HSS \citep[\textit{e.g.,}][]{2018garton}.

\section{Conclusions} \label{s:conclusion}

In this statistical study we investigated the evolution of $16$ long--living CHs between $2010$ and $2019$ and the associated evolution of the HSSs peak velocity at 1au. Our major findings can be summarized in the following:

\begin{enumerate}[i]
    \item We found that the general CH area evolution ($>80\%$ of the studied cases) exhibit a rough evolutionary profile showing a rise to a clear peak in the observed CH area followed by a decay.
    
    \item The strength and evolution of the photospheric magnetic flux and flux density enclosed in the projected CH boundary was found to be largely independent of the evolution of the CH area. This suggests that the mean signed flux density of CHs is not the main cause for the observed evolution in the CH area.
        
    \item Area growth and decay rates within CHs over their evolution seem to only vary over the solar cycle, with higher rates during lower solar activity. The calculated rates for the high solar activity match flux transport rates derived by \cite{1964Leighton}. Sudden large--scale changes in the area seem to be rare events.

    \item The rates at which the magnetic flux and flux density changes seems to be related to the activity of the Sun. We find higher rates during enhanced solar activity than during periods of low solar activity as approximated by the sunspot number.
    
    \item We find no correlation between the CH area and magnetic flux density changes rates. This hints that the rate at which the magnetic flux density of a CH evolves is independent from the CHs area evolution and and we suspect that this is caused by changes of the signed flux within the projected CH boundary (\textit{e.g.}, flux accumulation and cancellation due to flux emergence).

    \item It was shown that the well known CH area -- HSS also persists over each individual CH evolution with varying correlation coefficients and varying slopes of the linear regression line. 
\end{enumerate}

\begin{acknowledgements}
 The \textit{SDO} image data and the \textit{ACE} \textit{in--situ} data is available by courtesy of NASA and the respective science teams.  S.G.H., M.T. and A.M.V. acknowledge funding by the Austrian Space Applications Programme of the Austrian Research Promotion Agency FFG (859729, SWAMI). V.J. acknowledges support from University of Zagreb, Croatia, Erasmus+ program. M.D. acknowledges support by the Croatian Science Foundation under the project 7549 (MSOC). S.J.H. thanks the OEAD for supporting this research by a Mariett-Blau-fellowship. S.G.H. would like to thank A.H.-P. for his endless supply of coffee, great motivation and helpful discussions.
\end{acknowledgements}

\bibliographystyle{aa}

 \begin{appendix}
 \section{Snapshot of coronal holes under study}
\begin{figure*}[htbp]
   \centering
             \includegraphics[width=\textwidth]{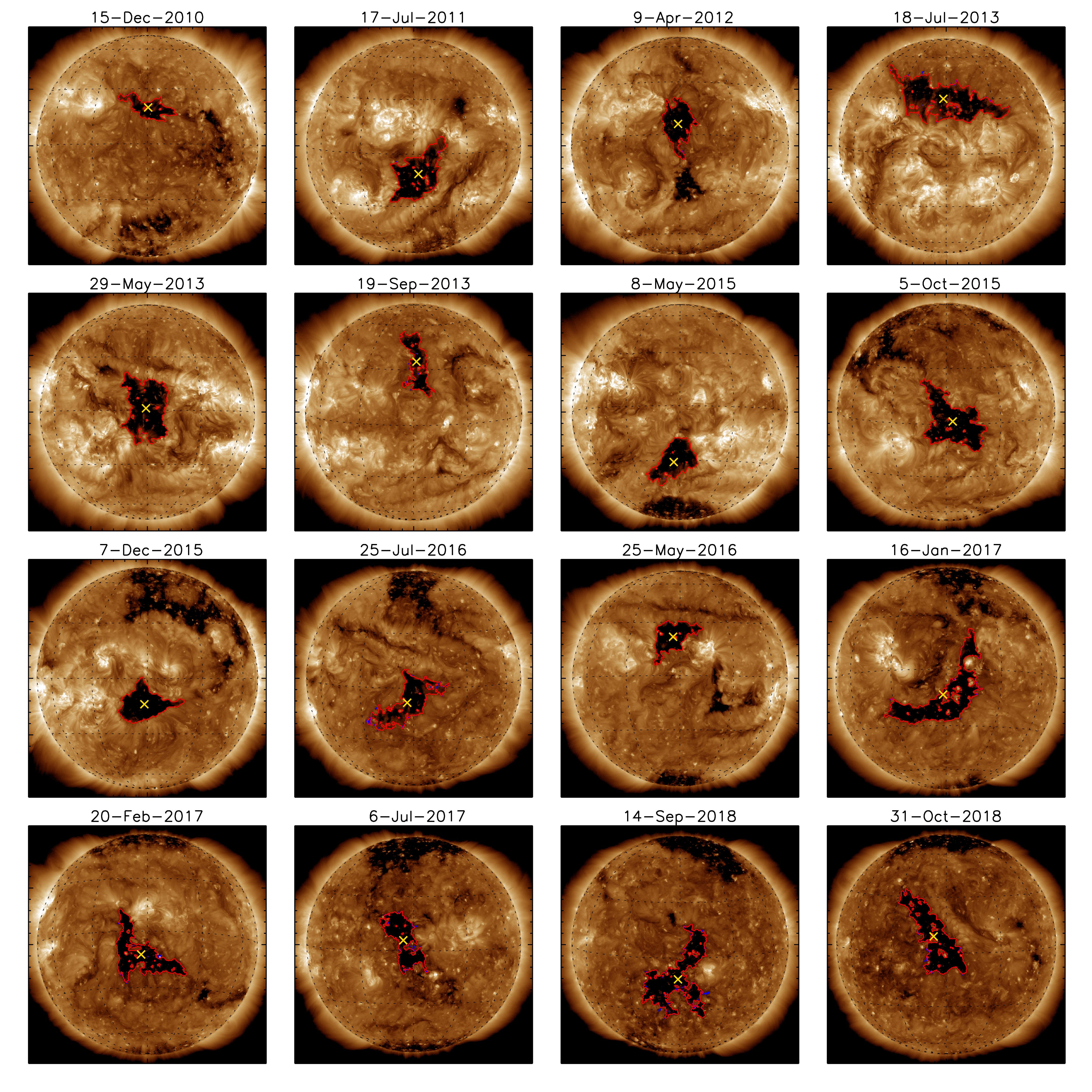}
       \caption{Example snapshot of CH under study. The extracted CH boundary is given by the red line and the center of mass is represented by the yellow x. The blue shaded areas represent the uncertainties as calculated by CATCH. }
          \label{fig:appendix1}
          
   \end{figure*}

Figure~\ref{fig:appendix1} shows a snapshot of every coronal hole analyzed in this study. The red line represents the extracted CH boundary using CATCH and the yellow x is the center of mass.

 \end{appendix}
 
  \end{document}